\documentclass{article}

\usepackage{spconf,amsmath,booktabs}
\usepackage{graphicx}
\usepackage{comment}
\usepackage{xcolor}
\usepackage{makecell}
\usepackage{multirow}
\usepackage{bm}
\usepackage{float}

\ninept
\title{Contrastive-Mixup Learning for Improved Speaker Verification}

\twoauthors
{Xin Zhang$^{\dagger}$\sthanks{Work done during an internship at Amazon.}}
{
Texas A\&M University\\
Dept.\ of Computer Science \& Engineering\\
College Station, TX, USA\\
\tt xin\_tamu@tamu.edu}
{\parbox{0.4\textwidth}{\centering
Minho Jin\sthanks{Equal contributions}, Roger Cheng, Ruirui Li, \\
Eunjung Han, Andreas Stolcke}}
{Amazon Alexa AI\\
Sunnyvale, CA, USA\\
\tt\hspace{-1cm}\{minhoj,rogcheng,ruirul\}@amazon.com\\
\tt \{cehan,stolcke\}@amazon.com}

\begin{document}

\maketitle

%

\begin{abstract}
This paper proposes a novel formulation of prototypical loss with mixup for speaker verification. Mixup is a simple yet efficient data augmentation technique that fabricates a weighted combination of random data point and label pairs for deep neural network training.
Mixup has attracted increasing attention due to its ability to improve robustness and generalization of deep neural networks. Although mixup has shown success in diverse domains, most applications have centered around closed-set classification tasks. In this work, we propose contrastive-mixup, a novel augmentation strategy that learns distinguishing representations based on a distance metric. During training, mixup operations generate convex interpolations of both inputs and virtual labels. Moreover, we have reformulated the prototypical loss function such that mixup is enabled on metric learning objectives. To demonstrate its generalization given limited training data, we conduct experiments by varying the number of available utterances from each speaker in the VoxCeleb database. Experimental results show that applying contrastive-mixup outperforms the existing baseline, reducing error rate by 16\% relatively, especially when the number of training utterances per speaker is limited. 
\end{abstract}

\section{Introduction}

Extensive research has been devoted to improving speaker recognition systems. The objective of speaker verification is to answer the question "Was the input spoken by an enrolled speaker?". The performance of a speaker verification system relies on access to sufficient and clean training data for supervised training \cite{Gudivada2017DataQC}. However, one of the challenges in training speaker verification models is the lack of large amounts of well-labelled training data. In other domains, researchers have developed various data augmentation techniques to overcome this bottleneck, enhancing the generalization of deep networks given limited data.
For example, data augmentation has been used in computer vision \cite{Perez2017TheEO}, natural language processing \cite{Feng2021ASO}, and semi-supervised learning \cite{Chaitanya2019SemiSupervisedAT}. 

Some well-known data augmentation approaches are applicable to speech recognition. Examples include adding noise, time stretching, pitch shift, and SpecAugment \cite{Wei_2020, Park_2019}. The common idea behind these approaches is to deform either the raw audio or the spectrogram with various operations, such as time and frequency masking. By applying these augmentations, embedding extractors learn to be more robust to variations and thus enable better generalization \cite{pmlr-v119-wu20g}.

In this paper, we propose contrastive mixup for training of a neural network's embedding extractor. Mixup is a regularization technique that trains the network with linear interpolations of input samples and corresponding interpolated labels. While the mixup technique has proven effective for closed-set classification tasks \cite{zhang2018mixup}, it is not clear how well it works for open-set applications such as speaker verification. Recent work has shed light on why mixup leads to improved robustness and generalization of the trained model from a theoretical perspective \cite{zhang2021does}. However, to the best of our knowledge, little work has been done to apply mixup to speaker verification systems. To fill this gap, we have developed contrastive mixup, a variant of the original mixup technique that is compatible with the training of speaker verification models. 

The key innovation of this paper is to demonstrate how to implement mixup for contrastive learning with metric learning objectives. We focus on speaker verification, which represents an open-set classification task. The speaker embeddings are extracted using a ResNet-based backbone model. To make a verification decision, the distance (cosine distance in this paper) is computed between the extracted embedding and the profile. The model is trained with batches consisting of a predefined number of utterances from a predefined number of speakers. During training, the prototype is calculated based on the original utterances from each individual speaker. Importantly, the query utterance is obtained by conducting linear interpolations of utterances from different speakers. Moreover, we choose angular prototypical loss to establish different baselines in speaker verification systems due to the results reported in previous work \cite{Chung_2018,heo2020clova}.

The prototypical networks are originally formulated for problems of few-shot learning, where each class can be represented and discriminated based on the mean of corresponding examples. One of the advantages of prototypical networks is that trained models can learn rare cases after being exposed to a small amount of prior information \cite{snell2017prototypical}. However, it still remains unclear how to integrate the mixup algorithm with prototypical loss because the original prototypical loss function solely relies on a distance metric between samples, not involving the use of labels. Consequently, we have to reformulate the prototypical loss function by taking advantage of label information such that the mixup operation can be incorporated in the metric learning objective.

\begin{figure*}[t]
    \centering
    \includegraphics[width=0.65\textwidth]{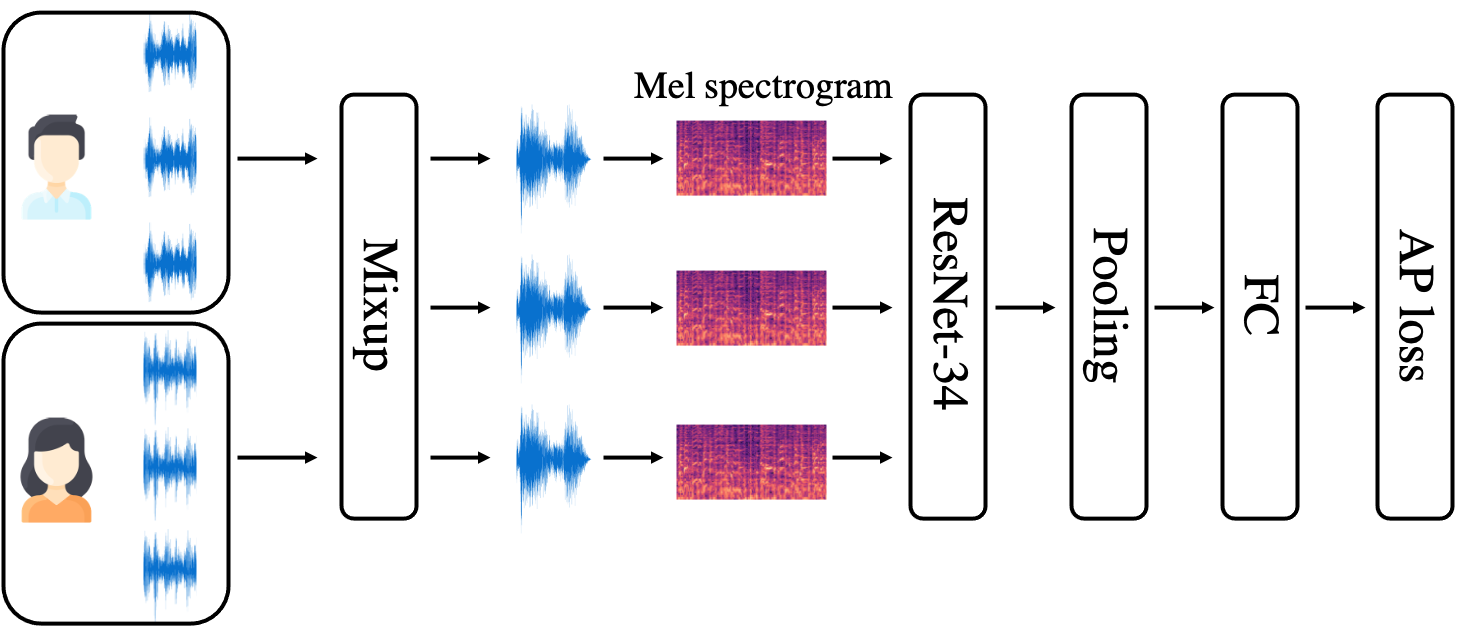}
    \caption{Schematic diagram of contrastive mixup}
    \label{fig:contrastive-mixup framework}
\end{figure*}

Furthermore, we conducted experiments to compare different types of implementations of the mixup algorithms, in both
time-domain speech samples and Mel spectrogram features. The results demonstrate that contrastive mixup is effective in improving speaker verification performance, especially when the number of utterances for each speaker is limited. Differently from existing augmentation approaches, such as adding noise \cite{snyder2015musan} and room response simulation \cite{doi:10.1121/10.0002489}, we believe contrastive mixup offers a novel and effective augmentation strategy that improves generalization while introducing negligible computational overhead.

The remainder of this paper is structured as follows: Section~\ref{sec:related_work} describes prior work related to mixup. Section~\ref{sec:contrastive-mixup-model} presents the proposed contrastive mixup, Section~\ref{sec:experiments} shows results and observations, and Section~\ref{sec:Conclusions} summarizes our findings.

\section{Related Work}
\label{sec:related_work}
Hongyi {\it et al.} \cite{zhang2018mixup} proposed the original version of the mixup algorithm by generating linear interpolations of data-label pairs. They empirically demonstrated that the linear behavior reduces the amount of undesirable oscillations when predicting outside the training examples. Recent works have shed light on the theoretical understanding of mixup regularization on deep neural networks. For example, \cite{thulasidasan2020mixup} shows that neural networks trained with mixup are significantly better calibrated and less prone to over-confident predictions on random noise data. \cite{carratino2020mixup} points out that random perturbation of the new interpretation of mixup leads to label smoothing and reduction of the Lipschitz constant of the estimator. In addition, several mixup variants have been proposed across a variety of tasks and models. For instance, feature-level interpolation and other types of transformations have been studied \cite{pmlr-v97-verma19a, summers2019improved, guo2018mixup}. \cite{yun2019cutmix} proposes the CutMix augmentation strategy, improving the performance of image classifier. 

From the perspective of training functions, the mixup method has been primarily utilized in conjunction with classification loss where virtual data points and virtual labels are fabricated by randomly interpolating original data points and ground-truth labels. Unlike classification tasks that aim to avoid misclassification, the goal of metric learning is to minimize the intra-class distances while expanding the inter-class distances. Metric learning objectives have been successfully employed for applications including speaker verification. For example, triplet loss \cite{Schroff_2015} and generalized end-to-end (GE2E) loss \cite{wan2020generalized} have shown remarkable performance for speaker verification when compared to conventional classification loss \cite{Chung_2018}. Therefore, it becomes an intriguing question how mixup augmentation might be combined with metric learning objectives to improve speaker verification performance.



\section{Contrastive mixup model}
\label{sec:contrastive-mixup-model}
Fig.~1 illustrates the overall architecture of contrastive mixup. Given a batch of utterances, utterances from different speakers are  randomly mixed to fabricate mixed inputs. Mixed utterances are fed into the backbone network, and finally a novel contrastive mixup loss function is applied.

\subsection{Angular Prototypical Loss}
The original prototypical network is designed for few-shot learning tasks \cite{snell2017prototypical} where good generalization can be achieved by training on a small number of samples in each class. Recent work demonstrates that metric learning objectives, such as generalized end-to-end and angular prototypical networks, outperform conventional classification objectives for speaker recognition. In particular, Chung {\it et al.} \cite{Chung_2018} presented a comprehensive study for metric learning in speaker recognition and conclude that angular prototypical loss outperforms state-of-the-art methods. In this paper, we choose angular prototypical (AP) loss to establish our baselines.

During training, suppose each batch contains $M$ utterances from $N$ different speakers. The embedding of each speaker is $x_{j,i}$, where $1 \leq j \leq N$ and $1 \leq i \leq M$. For the prototypical network, each batch consists of a support set $S$ and a query set $Q$. The $M$th utterance is set to be the query utterance. The centroid of speaker $j$ is computed:

\begin{equation}
    c_{j} = \frac{1}{M - 1} \sum_{m=1}^{M-1} \mathbf{x}_{j,m}
\end{equation}
Then, the distance between $\mathbf{x}_{j, M}$ and $\mathbf{c}_k$ are computed using trainable scale and bias coefficients, $w$ and $b$, as follows:

\begin{equation}
    \mathbf{S}_{j,k} = w \cos{(\mathbf{x}_{j, M}, \mathbf{c}_k)} + b
\end{equation}
Finally, the angular prototypical loss is computed using exponentiation as follows: 

\begin{equation}
    L_p = -\frac{1}{N} \sum_{j=1}^N \mathrm{log} \frac{e^{\mathbf{S}_{j,j}}} {\sum_{k=1}^N e^{\mathbf{S}_{j,k}}} \label{eq:L_p}
\end{equation}
As shown in (\ref{eq:L_p}), and as reported in \cite{chung2020defence}, angular prototypical loss outperforms the original prototypical networks and GE2E for speaker recognition tasks. As a result, we use AP loss as the baseline loss function for model training in our experiments. 

\subsection{Mixup}
Mixup training is based on the principle of vicinal risk minimization where the classifier is trained in the vicinity of each training sample \cite{zhang2018mixup}. Despite its simplicity, mixup has succeeded in a wide range of applications including computer vision \cite{zhang2018mixup}, natural language processing \cite{guo2020nonlinear}, and semi-supervised learning \cite{berthelot2019mixmatch}. In mixup, the input data and its associated label is modified as follows:
\begin{equation}
\begin{aligned}
    \overline{x}_{i}=\lambda x_{i} + (1-\lambda)x_{\mathrm{R}_i} \\
    \overline{y}_{i}=\lambda y_{i} + (1-\lambda)y_{\mathrm{R}_i} \label{eq:mixup}
\end{aligned}
\end{equation}
where $x_{i} (0 \le i \le B-1)$ is the $i$th data in the batch of size $B$, and $y_{i}$ is the one-hot encoding of the label. $\mathrm{R}_i$ is the $i$th randomly shuffled index.
For each training batch, the interpolation parameter $\lambda$ is sampled from a symmetric $Beta(\alpha, \alpha)$ distribution with $\alpha \in (0, \infty)$. Given fabricated input-label pairs, $(\overline{x}, \overline{y})$, the loss function for the classification task is computed as follows:
\begin{equation}
\begin{aligned}
    L_{mixup}(\overline{x}, y, y_{\mathrm{R}}, \lambda) = \lambda CE(\overline{x}, y) + (1-\lambda) CE(\overline{x}, y_{\mathrm{R}}),
    \label{eq:mixup_one}
\end{aligned}
\end{equation}
where CE denotes the cross-entropy loss. To apply mixup to AP loss, our first approach is to use cross entropy losses with respect to the ground truth and the shuffled labels as follows: 
\begin{eqnarray} 
    && L_{CE\_mixup} = \nonumber \\
    && -\frac{1}{N} \sum_{i=1}^N  \left [ \lambda \mathrm{log} \frac{e^{\overline{\mathbf{S}}_{i,i}}} {\sum_{k=1}^N e^{\overline{\mathbf{S}}_{i,k}}} + (1-\lambda)  \mathrm{log} \frac{e^{\overline{\mathbf{S}}_{i,\mathrm{R}_i}}} {\sum_{k=1}^N e^{\overline{\mathbf{S}}_{i,k}}} \right ]
    \label{eq:initial_loss}
\end{eqnarray}
by interpolating the losses with the ground truth and the shuffled labels. In the rest of this paper, (\ref{eq:initial_loss}) is referred to CE mixup. When applying mixup for raw speech signals, it is critical to normalize the volume when interpolating utterances from different speakers as it adjusts sound from diverse sources to the same volume level. 

\subsection{Contrastive-mixup Loss Function}

While mixup approach has been widely used for classification tasks, it remains a question how to implement mixup with metric learning objectives. Herein, we reformulate the AP loss by introducing a binary label function $d_{j,k}$:
\begin{equation} 
\begin{aligned}
    L_{c\_mixup} = -\frac{1}{N} \sum_{j=1}^N \mathrm{log} \frac{\sum_{k=1}^N d_{j,k} e^{\overline{\mathbf{S}}_{j,k}}}{\sum_{k=1}^N e^{\overline{\mathbf{S}}_{j,k}}}
    \label{eq:contrastive-mixup}
\end{aligned}
\end{equation}
where $0 \le d_{j,k} \le 1$. When $d_{j,k} = 1$ for $j=k$ and 0 otherwise, this becomes identical to (\ref{eq:L_p}). In this paper, we set $d_{j,k}$ as follows:
\begin{equation}
d_{j,k} = 
\begin{cases}
    \lambda,& \text{if } k = j\\    
    1-\lambda,& \text{if } k = \mathrm{R}_j\\
    0              & \text{otherwise}
\end{cases}
\end{equation}
In this scenario, the binary label $d_{j,k}$ can act as the virtual label for speaker/utterance pairs. When applying mixup on reformulated AP loss, the convex interpolation can be performed on both original utterances and binary labels.
We investigate the performance of contrastive mixup relative to the vanilla AP loss function.
\section{Experiments}
\label{sec:experiments}
\subsection{Data Description}
Our models are trained using the development set of VoxCeleb2 dataset \cite{chung2020defence}. VoxCeleb2 dataset contains more that 1 million utterances collected from nearly 6,000 speakers under controlled conditions. In order to obtain a fair comparison with previous results, the model evaluation is performed based on the test set of VoxCeleb1 \cite{Nagrani_2017}. 

\subsection{Training Details}
\begin{table}[!t]
    \small
    \centering
    \caption{Statistics on downsampled dataset with limited utterances per speaker}
    \addtolength{\tabcolsep}{1pt}
    \begin{tabular}{lccc}
    \toprule
    \makecell[l]{Number of\\ speakers} & \makecell[c]{Utterances per\\ speakers} & \makecell[c]{Total\\ utterances} & \makecell[c]{Percentage of\\ VoxCeleb2} \\
    \midrule
    \multirow{4}{*}{5994} & 2 & 11,988 & 1.0\% \\
    & 3 & 17,982 & 1.6\% \\
    & 5 & 29,970 & 2.6\% \\
    & 10 & 59,940 & 5.1\% \\
    \bottomrule
    \end{tabular}
    \label{tab:stat}
\end{table}
In our experiments, we employed PyTorch code \cite{chung2020defence} as a baseline. During training, 2s-long segments are randomly extracted from each utterance. The input feature, log Mel-spectrogram, was extracted every 10\,ms using a 25-ms window size. As shown in \cite{chung2020defence}, Fast ResNet34, which has a quarter-size channel compared to the original ResNet-34 \cite{he2015deep}, was used as a backbone model. The encoded output is aggregated using  self-attentive pooling (SAP) \cite{bhattacharya2017deep} to generate utterance-level representations. 

The models in this paper were trained using four NVIDIA V100 Tensor core GPUs with distributed configuration. We use the Adam optimizer with a learning rate of 0.001, decreasing by 5\% for every 10 epochs. Unless specified elsewhere, all models are trained for 500 epochs using a distributed configuration with four GPUs.

As indicated in \cite{chung2020defence}, large batch size leads to improved performance for metric learning methods due to the ability to sample hard negative samples within the batch. Therefore, we intentionally choose the largest batch size without exceeding the memory limits of the GPUs. For Quarter ResNet-34, the batch size is $400 \times 2$, namely the number of speakers $\times$ the number of utterances per speaker. For fair and reliable comparisons, all experiments are repeated three times, with mean and standard deviation calculated.

In our effort to investigate the effect of contrastive mixup, given that utterances per speaker are limited, we build a small training dataset by randomly sampling a certain number of utterances from each speaker. As shown in Table~\ref{tab:stat}, the reduced training datasets consists of 2, 3, 5, and 10 utterances per speaker, which is significantly smaller compared to the original VoxCeleb2. Then, we compare different training strategies using those down-sampled datasets.

The evaluation metric used in this work is equal error rate (EER) in \%, where false acceptance rate (FAR) and false rejection rate (FRR) are closest.
To demonstrate the effect of contrastive mixup across different training settings, we conduct experiments and compare the resulting EERs for the following cases:
\begin{enumerate}
\item original AP loss without augmentation and mixup;
\item original AP loss plus augmentation without mixup;
\item contrastive mixup without augmentation;
\item contrastive mixup plus augmentation.
\end{enumerate}
Here augmentation refers to noise addition and room impulse response (RIR) application based on the widely-used MUSAN corpus \cite{snyder2015musan}, which should be distinguished from the proposed contrastive mixup approach.
    
\subsection{Baseline and Evaluation Details}
We use the VoxCeleb1 \cite{Nagrani_2017} test set to evaluate model performance. Unlike the training stage, we sample ten 4-second temporal crops at regular intervals from each test segment. The similarities between all segment pairs across utterances are computed. The mean value of the similarities is calculated as the final score. A similar protocol can be found in \cite{Chung_2018}.

We focus on two critical aspects. First, how does contrastive mixup compare to the vanilla AP loss, and second, how does contrastive mixup behave with limited training data?
To this end, we train the models using different sizes of datasets. Note that all the other settings and hyperparameters are kept constant. 

In Tables~\ref{tab:contrastive_up}--\ref{tab:online_aug}, the baseline EER refers to training using original AP loss without augmentation. The mixup EER refers to training using the proposed contrastive-mixup loss without augmentation. The augmentation EER refers to training using original AP loss with augmentation \cite{snyder2015musan}. The mixup + augmentation EER refers to training using contrastive-mixup loss as well as augmentation. In addition, when employing contrastive-mixup loss, we fine-tune the mixing coefficient $\lambda$ to achieve the optimal performance.

\subsection{Results and Discussions}

In our preliminary analysis, we found that the performance of contrastive mixup is closely dependent on the choice of hyperparameter $\lambda$. Therefore, we have explored a range of $\lambda \sim Beta(\alpha, \alpha)$, where $\alpha$ equals 0.1, 0.2, 0.4, or 0.6, and compared the resulting EERs. Experimental results indicate that when training with the entire VoxCeleb2 dataset, setting $\alpha$ to be a smaller value (0.1) gives the best performance, whereas setting $\alpha$ to a larger value (e.g., 0.4) gives the smallest EER when training data is limited. A  previous study \cite{thulasidasan2020mixup} shows that relatively small values of $\alpha \in [0.1, 0.4]$ produce the best performance for classification, whereas a large value of $\alpha$ results in significant under-fitting. Our results agree well with previous observations in different domains \cite{zhang2018mixup, lee2020mix}. Also, the mixup is performed at the raw waveform level instead of the Mel-spectrum level based on our preliminary results, where the waveform level and the Mel-spectrum level give EERs of 2.11 $\pm$ 0.02 and 2.44 $\pm$ 0.06, respectively. Unless specified, the EER and its standard deviation in this paper were computed by repeating the experiments three times.

\begin{table}[!t]
    \small
    \centering
    \caption{Results with original AP loss ($L_{p}$) compared to two versions of mixup loss ($L_{CE\_mixup}$ (\ref{eq:initial_loss}) and $L_{c\_mixup}$ (\ref{eq:contrastive-mixup})).}
    \addtolength{\tabcolsep}{-2pt}
    \begin{tabular}{lccc}
    \toprule
        Type & \makecell[c]{Loss\\ function} & \makecell[c]{EER (\%)} & \makecell[c]{Relative\\ improvement} \\
    \midrule
    Baseline & Original AP \cite{chung2020defence} & 2.21 $\pm$ 0.03 & - \\
    \midrule
    \multirow{2}{*}{Mix-up} & 
    CE (\ref{eq:initial_loss}) & 2.19 $\pm$ 0.01 & +0.90\%\\
        & Contrastive (\ref{eq:contrastive-mixup}) & 2.11 $\pm$ 0.02 & +4.52\% \\
    \bottomrule
    \end{tabular}
    \label{tab:contrastive_up}
\end{table}

\begin{table}[!t]
    \small
    \centering
    \caption{Comparison of original AP loss ($L_{p}$) and contrastive-mixup loss ($L_{c\_mixup}$) on small datasets.}
    \addtolength{\tabcolsep}{1pt}
    \begin{tabular}{crrr}
    \toprule
    \makecell[l]{Utterances\\ per speaker} & \makecell[c]{Baseline\\ EER (\%)} & \makecell[c]{Mixup\\ EER (\%)} & \makecell[c]{Relative\\ improvement} \\
    \midrule
    2 & 14.80 $\pm$ 0.25 & 12.38 $\pm$ 0.24 & +16.3\% \\
    3 & 12.32 $\pm$ 0.33 & 10.53 $\pm$ 0.12 & +14.5\% \\
    5 & 9.43 $\pm$ 0.14 & 8.26 $\pm$ 0.07 & +12.4\% \\
    10 & 6.63 $\pm$ 0.08 & 6.05 $\pm$ 0.14 & +8.7\% \\
    \bottomrule
    \end{tabular}
    \label{tab:tab2}
    \label{tab:contrastive_up_limited_data}
\end{table}

Table~\ref{tab:contrastive_up} presents the EERs of baseline and two different versions (loss functions) of the mixup algorithm. Given the entire VoxCeleb2 training data set, we observed that mixup implementations improved with a modest gain of 4.52\%. Based on this result, in the follow-up experiments, we used contrastive mixup (\ref{eq:contrastive-mixup}).

Table~\ref{tab:tab2} compares the EERs with limited training data, where the number of utterances per speaker is limited to 2, 3, 5, or 10, to support our hypothesis that mixup improves generalization.  The gain with mixup was modest when the full VoxCeleb training data was used, as shown in Table~\ref{tab:contrastive_up}. As the number of utterances per speaker is reduced, we observed that contrastive mixup shows larger performance improvements relative to the baseline. When training on only two utterances per speaker, contrastive mixup can reduce the EER by up to 16.3\%.  This supports our hypothesis that mixup works as a generalization enhancement, especially with limited training data.

In Table~\ref{tab:online_aug}, contrastive mixup is applied on top of online augmentation. In these experiments, we fine-tune the mixing coefficient by choosing $Beta(0.6, 0.6)$ for 2 utterances, $Beta(0.2, 0.2)$ for 3 utterances, $Beta(0.4, 0.4)$ for 5 utterances, and $Beta(0.1, 0.1)$ for 10 utterances. While augmentation improves over the baseline, adding contrastive mixup on top of augmentation gives additional improvements, as shown in Table~\ref{tab:online_aug}. 

\begin{table}[!t]
    \small
    \centering
    \caption{Comparison of mixup on top of online augmentation}
    \addtolength{\tabcolsep}{-2pt}
    \begin{tabular}{crrr}
    \toprule
    \makecell[l]{Utterances\\ per speaker} & \makecell[c]{Augmentation\\ EER (\%)} & \makecell[c]{Mixup + Aug. \\ EER (\%)} & \makecell[c]{Relative\\ improvement} \\
    \midrule
    2 & 11.21 $\pm$ 0.28 & 10.75 $\pm$ 0.15 & +4.1\%\\
    3 & 9.15 $\pm$ 0.33 & 8.98 $\pm$ 0.21 & +1.8\% \\
    5 & 7.65 $\pm$ 0.12 & 7.55 $\pm$ 0.03 & +1.3\% \\
    10 & 5.86 $\pm$ 0.12 & 5.84 $\pm$ 0.07  & +0.3\%\\
    \bottomrule
    \end{tabular}
    \label{tab:online_aug}
\end{table}


As indicated by the evaluation results, while performance improvement is modest when training on the full VoxCeleb dataset, contrastive mixup consistently outperforms the baseline when the number of utterances per speaker is limited. The relative improvement between contrastive mixup and the baseline increases as the number of training utterance is reduced. Thus, our experimental results support the hypothesis that mixup boosts the generalization of neural network models.

\section{Conclusions}
\label{sec:Conclusions}
We propose contrastive mixup, a data augmentation strategy for contrastive representation learning of speaker verification systems. The key contribution of our work is a reformulation of the mixup loss function for metric learning objectives, specifically with angular prototypical loss. We show empirically that contrastive mixup consistently improves the performance of speaker verification models, especially when there are limited number of utterances per training speaker. Moreover, we observe that contrastive mixup can be applied on top of existing augmentation techniques to achieve further performance gains.

\vspace{-1em}
\section{Acknowledgments}
\vspace{-1em}
We would like to thank Oguz Elibol, Jasha Droppo and the Alexa SpeakerID team for their helpful feedback and discussions.

\normalsize


\bibliographystyle{IEEEtran}
\bibliography{mybib}

\end{document}